\documentstyle[onecolumn,epsfig]{mn}
\newif\ifAMStwofonts

\newcommand{\lsim}{\,\lower2truept\hbox{${<\atop\hbox{\raise4truept\hbox{$\sim$}}}$}\,}
\newcommand{\gsim}{\,\lower2truept\hbox{${>\atop\hbox{\raise4truept\hbox{$\sim$}}}$}\,}

\def\lsim{\,\lower2truept\hbox{${< \atop\hbox{\raise4truept\hbox{$\sim$}}}$}\,}
\def\gsim{\,\lower2truept\hbox{${> \atop\hbox{\raise4truept\hbox{$\sim$}}}$}\,}
\def\Ohat{{\widehat \Omega}}

\def\deg{\ifmmode^\circ \else$^\circ $\fi}    
\def\arcs{\ifmmode {'' }\else $'' $\fi}     
\def\arcm{\ifmmode {' }\else $' $\fi}     
\def\buildrel#1\over#2{\mathrel{\mathop{\null#2}\limits^{#1}}}
\def\mper{\ifmmode \buildrel m\over . \else $\buildrel m\over .$\fi}
\def\hper{\ifmmode \rlap.^{h}\else $\rlap{.}^h$\fi}
\def\sper{\ifmmode \rlap.^{s}\else $\rlap{.}^s$\fi}
\def\arcsper{\ifmmode \rlap.{' }\else $\rlap{.}' $\fi}
\def\arcmper{\ifmmode \rlap.{'' }\else $\rlap{.}'' $\fi}

\def\mincir{\ \raise -2.truept\hbox{\rlap{\hbox{$\sim$}}\raise5.truept	
\hbox{$<$}\ }}								%
\def\magcir{\ \raise -2.truept\hbox{\rlap{\hbox{$\sim$}}\raise5.truept	%
\hbox{$>$}\ }}								%

\title[Cosmological implications of future CMB spectrum measures at
long wavelengths]
{What can we learn on the thermal history of the Universe from 
future CMB spectrum measures at long wavelengths?}

\author[C. Burigana and R. Salvaterra]
{C. Burigana$^1$ and
R. Salvaterra$^2$\\ 
$^1$IASF/CNR, Istituto di Astrofisica Spaziale e Fisica Cosmica,
Sezione di Bologna, \\
Consiglio Nazionale delle Ricerche, 
Via Gobetti 101, I-40129 Bologna, Italy \\
$^2$SISSA/ISAS, Astrophysics Sector, Via Beirut, 4, I-34014 Trieste, Italy}

\date{Submitted to MNRAS, 29 December 2002.}

\pagerange{\pageref{firstpage}--\pageref{lastpage}}
\pubyear{2002}

\begin{document}

\maketitle

\label{firstpage}
\footnotetext{E-mail: burigana@bo.iasf.cnr.it ; salvater@sissa.it}
\begin{abstract}
We analyse the implications of future observations of the CMB 
absolute temperature at centimeter and decimeter wavelengths, 
where both ground, balloon and 
space experiments are currently under study to complement 
the accurate COBE/FIRAS data available at $\lambda \lsim 1$~cm.

Our analysis shows that forthcoming  
ground and balloon measures will allow a better understanding
of free-free distortions but 
will not be able to significantly improve 
the constraints already provided by the FIRAS data
on the possible energy exchanges in the primeval
plasma.
The same holds even for observations
with sensitivities up to $\sim 10$ times better than those 
of forthcoming experiments.

Thus, we have studied the impact of very high quality data, such those
in principle achievable with a space experiment like
DIMES planned to measure the CMB absolute 
temperature at $0.5 \lsim \lambda \sim 15$~cm with a sensitivity of
$\sim 0.1$~mK,
close to that of FIRAS.
We have demonstrated that such high quality data would 
improve by a factor $\sim 50$ 
the FIRAS results on the fractional energy exchanges,
$\Delta\epsilon/\epsilon_i$,
associated to dissipation processes 
possibly occurred in a wide range of cosmic epochs,
at intermediate and high redshifts ($y_h \gsim 1$),
and that the energy dissipation 
epoch could be also significantly constrained.
By jointly considering two dissipation processes occurring 
at different epochs, we demonstrated that 
the sensitivity and frequency coverage of a DIMES-like 
experiment would allow to accurately recover the epoch and 
the amount of energy possibly injected in the radiation field
at early and intermediate epochs even in presence of
a possible late distortion, while 
the constraints on the 
energy possibly dissipated at late epochs 
can be improved by a factor $\simeq 2$.
In addition, such measures can provide an independent
and very accurate cross-check of FIRAS calibration.

Finally, 
a DIMES-like experiment will be able to provide indicative 
independent estimates of the baryon density: the 
product $\Omega_b H_0^2$ can be recovered 
within a factor $\sim 2 - 5$ even in the case of (very small) early
distortions with $\Delta\epsilon/\epsilon_i \sim (5 - 2) \times 10^{-6}$.
On the other hand, for $\Omega_b (H_0/50)^2 \lsim 0.2$,
an independent baryon density 
determination with an accuracy
at $\sim$ per cent level, comparable to that achievable
with CMB anisotropy experiments, 
would require an accuracy of $\sim 1$~mK or better in the measure 
of possible early distortions but 
up to a wavelength from $\sim$~few~$\times$~dm to $\sim 7$~dm, according
to the baryon density value.

\end{abstract}


\section{Introduction}

As widely discussed in many papers, the spectrum of the 
Cosmic Microwave Background (CMB) carries unique informations on physical
processes occurring during early cosmic epochs
(see e.g. Danese \& Burigana 1993 and references therein).
The comparison between models of CMB spectral distortions
and CMB absolute temperature measures can constrain the
physical parameters of the considered dissipation processes.
We recently discussed (Salvaterra \& Burigana 2002) 
the implications of the current CMB spectrum data
by jointly considering distortions generated in a wide range of 
early or intermediate cosmic epochs and at late cosmic epochs. 

Various CMB spectrum experiments at long wavelengths,
$\lambda \gsim 1$~cm, 
are ongoing and planned for the future in order
to improve the still quite poor accuracy of the data 
in this spectral region, where the maximum deviations from
a pure blackbody spectrum are expected in the case of 
dissipation processes occurred at early and intermediate
epochs.

In this work we jointly consider the data from the FIRAS instrument 
aboard the COBE satellite and simulated sets
of CMB spectrum observations at wavelengths larger than 1~cm
with the sensitivities expected from future experiments
in order to discuss their impact for the recovery of the 
thermal history of the Universe.

In section~2
we briefly summarize the general properties of the
CMB spectral distortions and the main physical informations 
that can be derived from the comparison with the observations.
In section~3 we briefly discuss the performances of current and future 
CMB spectrum observations at long wavelengths and describe the generation
of the simulated observations used in this work.
The implications of observations with sensitivities typical of 
forthcoming and future ground and balloon experiments are presented in
section~4, while
in section~5 we extensively discuss the implications of experiments
at long wavelengths with a sensitivity comparable to that of COBE/FIRAS,
as foreseen for a space experiment, DIMES, proposed to the NASA in the 1995
and designed to measure the CMB absolute temperature 
at $\simeq 0.5 - 15$~cm with a sensitivity of $\sim 0.1$~mK (Kogut 1996).
In section~6 we present a detailed discussion of the 
capabilities of future CMB spectrum observations to discriminate 
between the FIRAS
calibration by the COBE/FIRAS team (referred as here ``standard''
calibration; see Fixsen et al. 1994, 1996, 
Mather et al. 1999, and references therein)
and that proposed by Battistelli, Fulcoli \& Macculi 2000.
The possibility to improve our knowledge of the free-free distortions
is considered in section~7, while section~8 is devoted to identify 
the experimental sensitivity requirements for an accurate baryon density 
evaluation through the detection of possible long wavelength distortions.
Finally, we draw our main conclusions in section~9.

\section{Theoretical framework}

The CMB spectrum emerges from the thermalization redshift, 
$z_{therm} \sim 10^6 - 10^7$, 
with a shape very close to a Planckian one, 
owing to the strict coupling between radiation and matter through
Compton scattering and photon production/absorption processes, 
radiative Compton and Bremsstrahlung,
which were extremely efficient at early times 
and able to re-establish a blackbody (BB) spectrum 
from a perturbed one
on timescales much shorter than the expansion time (see e.g. 
Danese \& De~Zotti 1977).
The value of $z_{therm}$ (Burigana, Danese \& De~Zotti 1991a)
depends on the baryon density (in units of the critical density),
$\Omega_b$, 
and the Hubble constant, $H_0$, through the product 
$\Ohat_b =\Omega_b (H_{0}/50)^2$ ($H_0$ expressed in Km/s/Mpc). 

On the other hand, physical processes occurring at redshifts $z < z_{therm}$ 
may lead imprints on the CMB spectrum.


The timescale for the achievement of
kinetic equilibrium between radiation and matter
(i.e. the relaxation time for the photon spectrum), $t_C$, is
\begin{equation}
t_C=t_{\gamma e} {m c^{2}\over {kT_e}} \simeq 4.5 \times 10^{28} 
\left( T_{0}/2.7\, K \right)^{-1} \phi^{-1} \Ohat_b^{-1}
\left(1+z \right)^{-4} \sec \, ,
\end{equation}
where $t_{\gamma e}= 1/(n_e \sigma _T c)$ is the photon--electron collision
time, $\phi = (T_e/T_r)$, $T_e$ being the electron temperature and
$T_r=T_{0}(1+z)$;
$kT_e/mc^2$ is the mean fractional change of photon energy in a scattering
of cool photons off hot electrons, i.e. $T_e \gg T_r$;
$T_0$ is the present radiation temperature related
to the present radiation energy density by $\epsilon _{r0}=aT_0^4$;
a primordial helium abundance of 25\% by mass is here assumed.

It is useful to introduce the dimensionless time variable $y_e(z)$ defined by
\begin{equation}
y_e(z) = \int^{t_0}_{t} {dt \over t_C}
=\int^{1+z}_{1} {d(1+z) \over 1+z} {t_{exp}\over t_C} \, ,
\end{equation}
where $t_0$ is the present time and
$t_{exp}$ is the expansion time given
by
%
\begin{equation}
t_{exp} \simeq   6.3\times 10^{19} \left({T_0 \over 2.7\, K}\right)^{-2}
(1+z)^{-3/2} \left[\kappa (1+z) + (1+z_{eq})
-\left({\Omega_ {nr} -1 \over \Omega_ {nr}}\right)
\left({1+z_{eq} \over 1+z}\right) \right]^{-1/2}
\sec \, ,
\end{equation}
$z_{eq} = 1.0\times 10^4 (T_{0}/2.7\, K)^{-4}\Ohat _{nr}$
being the redshift of
equal non relativistic matter and photon energy densities
($\Omega _{nr}$ is the density of non relativistic matter in units of critical
density); $\kappa = 1 + N_\nu (7/8)
(4/11)^{4/3}$, $N_\nu$ being the number of relativistic, 2--component,
neutrino species (for 3 species of massless neutrinos, $\kappa \simeq 1.68$),
takes into account the
contribution of relativistic neutrinos to the dynamics of the
Universe\footnote{Strictly speaking the present ratio of neutrino to
photon energy densities, and hence the value of $\kappa$, is itself a
function of the amount of energy dissipated. The effect, however,
is never very important and is negligible
for very small distortions.}.

Burigana, De~Zotti \& Danese 1991b have reported on
numerical solutions of the Kompaneets equation (Kompaneets 1956)
for a wide range of values of the relevant parameters
%
%
and accurate analytical representations of these numerical solutions,
suggested in part by the general properties of the Kompaneets 
equation and by its well known asymptotic solutions,
have been found (Burigana, De~Zotti \& Danese 1995).

The CMB distorted spectra depend on at least
three main parameters: the fractional amount of energy exchanged between
matter and radiation, $\Delta\epsilon / \epsilon_i$,
$\epsilon _i$ being the radiation energy density before the energy injection,
the redshift $z_h$ at which the heating occurs, and the
baryon density $\Ohat_b$.
The photon occupation number can be then expressed in the form
\begin{equation}
\eta = \eta (x; \Delta\epsilon / \epsilon_i, y_h, \Ohat_b) \, ,
\end{equation}
where $x$ is the dimensionless frequency $x = h\nu/kT_{0}$
($\nu$ being the present frequency),
and $y_h \equiv y_e(z_h)$ characterizes the epoch when the energy dissipation
occurred, $z_h$ being the corresponding redshift
(we will refer to $y_h \equiv y_e(z_h)$ computed assuming $\phi=1$, so that the epoch 
considered for the energy dissipation does not depend on the 
amount of released energy).
The continuous behaviour of the distorted spectral shape with $y_h$ can be
in principle used also to search for constraints on the 
epoch of the energy exchange.
Of course, by combining the approximations describing the distorted spectrum
at early and intermediate epochs with the Comptonization distortion expression
describing late distortions, it is possible to jointly treat
two heating processes (see Burigana et al. 1995 and Salvaterra \&
Burigana 2002 and references therein for a more exhaustive discussion).

In this work
the measures of the CMB absolute temperature 
are compared with the above models
of distorted spectra for one or two heating processes
by using a standard $\chi^2$ analysis. 

We determine the limits on the amount of energy possibly injected 
in the cosmic background at arbitrary primordial epochs corresponding to a
redshift $z_h$ (or equivalently to $y_h$).
This topic has been discussed in several works
(see e.g. Burigana et al. 1991b,
Nordberg \& Smoot 1998, Salvaterra \& Burigana 2002). 
As in Salvaterra \& Burigana 2002, we improve here the
previous methods of analysis
by investigating the possibility of properly combining
FIRAS data with longer wavelength measures
with the sensitivities expected for forthcoming 
and future experiments
and by refining the method
of comparison with the theoretical models. 
We will consider the recent improvement in the
calibration of the FIRAS data, that sets the CMB scale temperature at
$2.725\pm0.002$~K at 95 per cent confidence level (CL) (Mather et al. 1999). 
We do not consider
the effect on the estimate of the amount of energy injected in the CMB
at a given epoch introduced by the calibration uncertainty of FIRAS 
scale temperature
when FIRAS data are treated jointly to longer wavelength measures, since the 
analysis of Salvaterra \& Burigana 2002 shows that it introduces only
minor effects.

Then, we study the combined effect of two different heating processes
that may have distorted the CMB spectrum at different epochs.
This hypothesis has been also taken into account by 
Nordberg \& Smoot 1998, who fit
the observed data with a spectrum distorted by a single heating at $y_h=5$,
a second one at $y_h\ll 1$ and by free-free emission, obtaining limits on the
parameters that describe these processes.
As in Salvaterra \& Burigana 2002, we extend their analysis by
considering the full range of epochs for the early and 
intermediate energy injection
process, by taking advantage of the analytical representation
of spectral distortions at intermediate redshifts (Burigana et~al. 1995).
Since the relationship between free-free distortion and Comptonization
distortion is highly model dependent,
being related to the details of the thermal history at late
epochs (Danese \& Burigana 1993, Burigana et al. 1995),
and can not be simply represented by integral parameters,
we avoid a combined analysis of free-free distortions 
and other kinds of spectral distortions
and separately discuss the implications of future,  
more accurate long wavelength measures on free-free distortions.

It is also possible to extend the limits 
on $\Delta\epsilon/\epsilon_i$ for heatings
occurred at $z_h>z_1$, where $z_1$ is the redshift corresponding
to $y_h = 5$, when the Compton
scattering was able to restore the kinetic equilibrium between matter and
radiation on timescales much shorter than the expansion time
and the evolution on the CMB spectrum can be easily studied
by replacing the full Kompaneets equation with the differential
equations for the evolution of the electron temperature 
and the chemical potential.
This study can be performed by using the
simple analytical expressions by Burigana et al. 1991b
instead of numerical solutions.

A recent analysis of the limits on the amount of
the energy possibly injected in the cosmic background 
from the currently available data is reported in 
Salvaterra \& Burigana 2002.
In particular, they found that the measures at 
$\lambda \gsim 1$~cm do not significantly 
contribute to these constraints because of their poor 
sensitivity compared to that 
of FIRAS. New and more accurate 
measurements are also needed in this range, which is particular 
sensitive to early energy injection processes.
In fact, the current constraints on $\Delta\epsilon/\epsilon_i$
at $y_h \sim 5$ are a factor $\sim 2$ less stringent than those at 
$y_h$ less than $\sim 0.1$, because of the frequency coverage of FIRAS,
which mainly set the current constraints at the all cosmic epochs.
Thus, we are interested to investigate the role of future ground, balloon
and space experiments at $\lambda \gsim 1$~cm 
jointed  to the FIRAS measures at $\lambda \lsim 1$~cm.

To evaluate the scientific impact represented by 
the future experiment improvements, we
create different data sets 
simulating the observation of a not distorted spectrum both 
from ground and balloon experiments and from a space experiment 
like DIMES 
through the method described in section~3.1.
For a DIMES-like experiment, we 
also explore the possibility of the observation of distorted spectra for
different amounts of the energy injected in the radiation field 
and for different cosmic epochs.

Each data set will be then compared to models of distorted spectra by using 
the method described in Salvaterra \& Burigana 2002
(see also Burigana \& Salvaterra 2000 for the details of the code)
to recover the value of 
$\Delta\epsilon/\epsilon_i$ or constraints on it, the heating epoch, $y_h$,
the free-free distortion parameter $y_B$,
and the combination, $\Ohat_b$,
of the baryon density and the Hubble constant.


For simplicity, we restrict 
to the case of a baryon density
$\Ohat_b = 0.05$ our analysis 
of the implications for the thermal history of the Universe,
but the method can be simply applied to different 
values of $\Ohat_b$.
In presence of an early distortion,
$\Ohat_b$ could be in principle
measured by CMB spectrum observations
at long wavelengths 
provided that they have the required sensitivity 
about the minimum of the CMB absolute temperature
(see section~8).

\section{Future experiments}

The CMB spectrum experiments currently under study are dedicated
to improve our knowledge at
wavelengths longer than those covered by FIRAS. At centimeter and 
decimeter wavelengths, the available measures typically show large error 
bars although some
experiments are rather accurate (i.e., the measure of Staggs
et~al. 1996 at $\simeq 2.8$~cm). Very accurate data at long wavelengths 
could give a significant improvement to our knowledge of physical processes
in the primeval plasma, particularly at high redshifts.
These projects regard measurements from ground, balloon and space.
As representative cases, and without the ambition to cover the whole 
set of planned experiments, we briefly 
refer here to the ground experiment TRIS
at very long wavelengths and to the 
DIMES experiment from space 
(Kogut 1996)
designed to reach an accuracy close to that of FIRAS 
up to $\lambda \simeq 15$~cm. 

TRIS\footnote{http://sunradio.uni.mi.astro.it/grupporadio/tris/index.html}
is a set of total power radiometers designed to measure the absolute
temperature of the CMB at three frequencies: 0.6, 0.82 and 2.5 GHz.

At these wavelengths ($12 - 50$~cm)
the measurements are difficult because 
the CMB signal is comparable to other components of the antenna 
temperature: Galactic background, unresolved extra-galactic sources,
sidelobes pickup and atmospheric emission. To improve the experimental 
situation, TRIS will make absolute maps of large areas of the sky at the 
three frequencies, to disentangle the various components of the celestial
signal; all the lossy parts of the
antenna front ends of the receivers will be 
cooled down liquid helium temperature, 
to reduce the thermal noise of these 
components; the receiver temperatures will be very carefully stabilized 
to reduce drifts and gain variations.
The TRIS expected sensitivity is of about 200~mK at the three frequencies.

DIMES\footnote{http://map.gsfc.nasa.gov/DIMES/index.html} (Diffuse 
Microwave Emission Survey) is a space mission submitted to the NASA in
1995, designed to measure very accurately the CMB spectrum at wavelengths
in the range $\simeq 0.5 - 15$~cm (Kogut 1996).

DIMES will compare the spectrum of each 10 degree pixel on the sky to a
precisely known blackbody to precision of $\sim 0.1$~mK,
close to that of FIRAS ($\simeq 0.02 - 0.2$~mK). 
The set of receivers is given
from cryogenic radiometers operating at six frequency bands 
about 2, 4, 6, 10 e 90~GHz using a single external blackbody 
calibration target common to all channels. 
In each channel, a cryogenic radiometer
switched for gain stability between an internal reference load and an
antenna with 10 degree beam width, will measure the signal change as the 
antenna alternately views the sky and an external blackbody calibration 
target. The target temperature will be adjusted to match the sky signal 
in the lowest frequency band, allowing the absolute temperature to be
read off from the target thermometry with minimal corrections for
the instrumental signature. With its temperature held constant, the 
target will rapidly move over the higher-frequencies antenna apertures,
effectively comparing the spectrum of diffuse emission from the sky 
to a precise blackbody. By comparing each channel to the same target, 
uncertainties in the target emission cancel so that deviations from a 
blackbody spectral shape may be determined much more precisely than the 
absolute temperature.
The DIMES design is driven by the
need to reduce or eliminate systematic errors from instrumental artifacts.
The instrument emission will be cooled to 2.7~K, whereas the calibration 
uncertainty will be minimized by using a single calibration target, common 
to all channels. The atmospheric emission will be observed from low Earth
orbit and the multiple channels measurements will minimize the 
foreground emission problems.
The DIMES sensitivity represents an
improvement by a factor better than 300 with respect to 
previous measurements at cm wavelengths. 

\subsection{Generation of simulated data sets} 

We collect different data sets, 
simulating measurements of different CMB spectra, distorted or not, at the 
frequency ranges of the considered experiments. 
We add to these simulated data 
the FIRAS data at
higher frequencies according to the most recent calibration
of the temperature scale at 2.725 K
(Mather et~al. 1999).

For the cases of distorted spectra
we calculate the theoretical temperature of the CMB spectrum at the
wavelengths of the new experiments
as discussed in the previous section.
Of course, 
the thermodynamic temperature
held obviously constant at all the frequencies
for the case of a non distorted spectrum. 

The theoretical temperatures
are then fouled to simulate real measurements affected by instrumental noise.
The simulated temperature $T_{obs}$ at the frequency $\nu$ is
given by 

\begin{equation}
T_{obs}(\nu)=T_{teor}(\nu)+n(\nu)\times \mbox{err}(\nu) \, ,
\end{equation}

\noindent
where $T_{teor}(\nu)$ is the theoretical temperature at the frequency $\nu$
and err($\nu$) is the expected rms error (at 1~$\sigma$) 
of the experiment at this frequency.
The numbers $n$ are a set of random numbers generated according to a Gaussian
distribution with null mean value and unit variance with the routine
GASDEV by Press et al. 1992 (\S 7).

\section{Implications of future ground and balloon experiments}\label{sec:terra}

We analyse here the impact of possible future observations 
from ground and balloon in the case of 
a not distorted spectrum at the temperature $T_0=2.725$~K. 
The results are 
thus comparable to those obtained with the FIRAS data alone 
(see e.g. Salvaterra \& Burigana 2002). 

\subsection{Simulated data sets}\label{dati_terra}

To build the first simulated data set (G\&B1-BB), 
we split the region from 1 to 80 cm in three
ranges and associate different values of sensitivity to each 
range according to the analysis of the main problems affecting
the available observations in different spectral 
regions (e.g. Salvaterra \& Burigana 2000).
 
\begin{enumerate}
\item $1 - 4$ cm. In this range the measurements of Staggs et~al. 1996 
show an uncertainty of $\sim 40$~mK. 
Thus, quite accurate measures could be carried out in this range. 
We choose to associate to the future experiments at these wavelengths
an improved typical sensitivity of 10~mK.
\item $4 - 9$~cm. Ground experiments in this range show error bars 
of about $50 - 70$~mK. Progresses 
could be reached by improving the accuracy of the subtraction
of the atmospheric contribution which dominates the final error 
at these wavelengths. Thus, we choose to associate to the data 
in this region a typical sensitivity of 40~mK;
\item $10 - 80$~cm. Observations in this range are still quite difficult, 
the typical sensitivities being 
between 200~mK for measures at 10 cm 
and 1.5~K for those at longer wavelengths. The expected sensitivity
of the TRIS experiment (see section~3)
is of $\sim 200$~mK. Thus, we choose to associate to future experiments 
in this range a typical sensitivity of 200~mK.
\end{enumerate}

The frequencies of the experiments 
at $\lambda \gsim 1$~cm of the two last decades
(see e.g. Table~1 of Salvaterra \& Burigana 2002),
where suitable observation windows should exist 
and the presence of man made interferences
should be not a concern,
have been adopted in the generation of simulated observations.

Finally, we complete this data set by adding the FIRAS measures 
calibrated at 2.725~K according to Mather et al. 1999 to the above
simulated data.

A second data set (G\&B2-BB) is built as before but by improving by a factor
10 the sensitivity associated to each of the above three frequency range
in order to evaluate the impact of 
highly optimistic future progresses
of ground and ballon experiments.

\subsection{Fits to simulated data}\label{fit_terra}

The results of the fits to the 
simulated data G\&B1-BB and G\&B2-BB 
jointed to FIRAS data are shown in Fig.~1
[for graphic purposes, we report in the plots the exact value
of $y_h$ and the power-law approximation
$z_h(y_h) \simeq 4.94 \time 10^4 y_h^{0.477} \Ohat_b^{-0.473}$
(Burigana et al. 1991b) for the redshift].
As evident, realistic improvements of future experiments from
ground and balloon do not significantly change the FIRAS limits.
Even under much more optimistic experimental conditions, able to decrease
the errors by a factor 10, 
the situation can not substantially improve, being 
the limits on 
$\Delta \epsilon /\epsilon _i$ 
obtained in this case 
only just more stringent than those based on FIRAS data alone.

We then conclude that, unfortunately,  
observations of the CMB absolute temperature
with sensitivity levels typical of 
future ground and balloon experiments do not seem able 
to improve the limits on the amount of energy injected in the
cosmic radiation field inferred on the basis of the currently available 
measures.

\begin{figure*}
\epsfig{figure=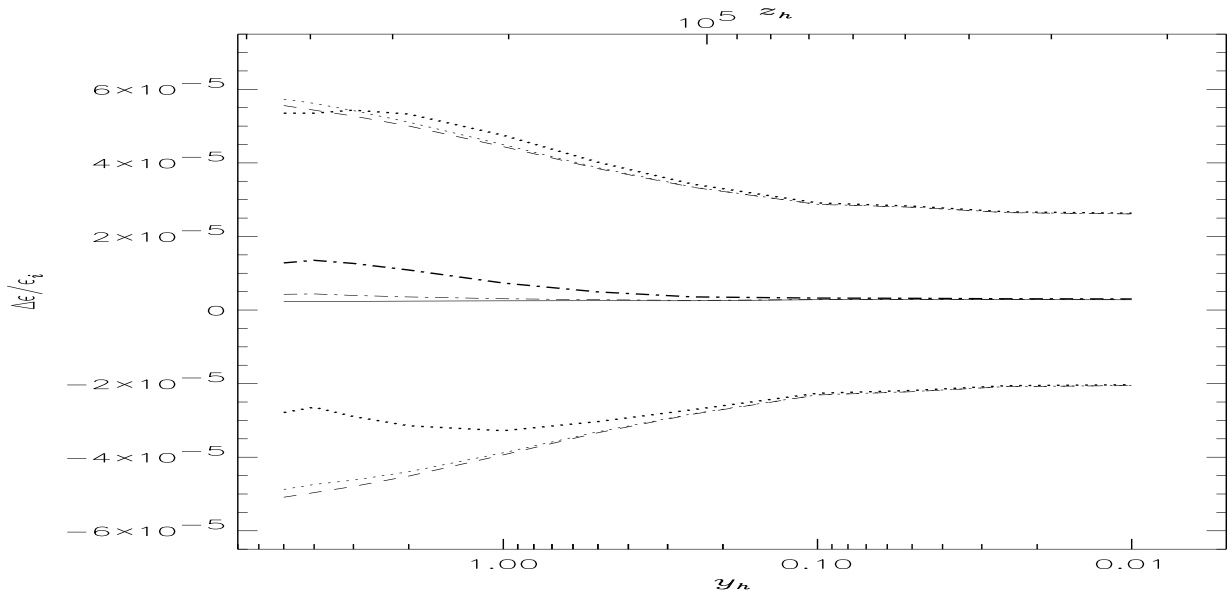,height=8cm,width=17cm}
\caption{Comparison between the constraints (at 95 per cent CL) on the energy exchanges
derived at different cosmic times from different data sets by considering
the case of a single dissipation process. 
The different lines refer to: 
FIRAS data calibrated according to Mather et al. 1999
(solid line: best fit; dashed lines: upper and lower limits); 
FIRAS data jointed to the 
data set G\&B1-BB (dot-dashes: best fit; dots: upper and lower limits);
FIRAS data jointed to the 
data set G\&B2-BB (thick dot-dashes: best fit; 
thick dots: upper and lower limits ).
}
\end{figure*}

\section{Implications of a DIMES-like experiment}\label{sec:dimes}

\subsection{Simulated data sets}\label{dati_dimes}

We generate a set (D-BB) of simulated data in the case of a blackbody
spectrum at a temperature of 2.725~K 
in order to evaluate the capability of
an experiment with a sensitivity comparable to that 
expect for DIMES to improve the constraints on the amount of the 
energy injected in the cosmic radiation field.
The analysis of this case is in fact directly 
comparable with the results obtained from the fit to the
FIRAS data alone.

Then, we build up other data sets representing the observations 
of CMB spectra distorted by
energy injections at different cosmic epochs 
in order to investigate the possibility of a DIMES-like experiments 
to firmly determine the presence of spectral distortions. 
We consider processes occurring at a wide range of cosmic epochs, represented
by the dimensionless time $y_h =$ 
 5, 4, 3, 2, 1, 0.5, 0.25, 0.1, 0.05, 0.025, 0.01, and 
$y_h\ll1$.
We consider four representative values of fractional injected energy:
$\Delta\epsilon/\epsilon_i = 2\times10^{-5}$, 
a value not much below the upper FIRAS limits; 
$\Delta \epsilon /\epsilon _i = 2\times10^{-4}$, 
well above the FIRAS upper limit (see section~6); 
$\Delta \epsilon /\epsilon _i = 5\times10^{-6}$ and $2\times10^{-6}$, 
two values well below the FIRAS upper limit,
to test the chances 
to detect very small distortions
with a DIMES-like experiment.

As a further representative case, we simulate the observation 
of a spectrum distorted by two heating processes occurring at
different epochs, 
the first at $y_h=5$ and the second at 
$y_h\ll1$, both characterized 
by $\Delta\epsilon/\epsilon_i=5\times10^{-6}$.

All these distorted spectra are computed by setting $\Omega_b=0.05$ 
and $H_0=50$~Km/s/Mpc.

As a variance with respect the previous section, we choose here the frequencies 
of the simulated observations by adopting 
the five frequency channels of the DIMES experiment.

As in previous section, we complete these data set by adding the 
FIRAS measures calibrated at 2.725~K.
 
\subsection{Fits to simulated data: analysis of a single dissipation
process}

\subsubsection{Non distorted spectrum}

We fit the simulated data D-BB with a spectrum distorted by an energy
injection at different values of $y_h$ in order to recover the value of 
$\Delta\epsilon/\epsilon_i$, expected to be null, and the limits on it.
The fit results are reported in Fig.~2.

It is evident how future data at this sensitivity level will allow a
strong improvement of the 
limits obtained with the FIRAS data alone.
The recovered best-fit value of 
$\Delta\epsilon/\epsilon_i$ is always compatible with 
the absence of distortions within the limits at 95 per cent CL. 
For heating processes at
low $z$ ($y_h= 0.1-0.01$) the fit is substantially dominated by
the FIRAS data and the lower and the upper limits on 
$\Delta\epsilon/\epsilon_i$ are still $\sim 2\times10^{-5}$.
On the contrary, for early distortions ($y_h \gsim 1$) 
the low frequency measures of a DIMES-like experiment will allow to
improve the FIRAS constraints by a factor $\sim 10-50$, the proper value
increasing with the considered dissipation redshift.
We conclude that measures from an instrument like DIMES 
could represent a very good complement to the FIRAS data.

In the next sections we will analyse in detail the capability of a DIMES-like 
experiment to determine the presence of spectral distortions possibly
present in the CMB spectrum.

\begin{figure*}
\epsfig{figure=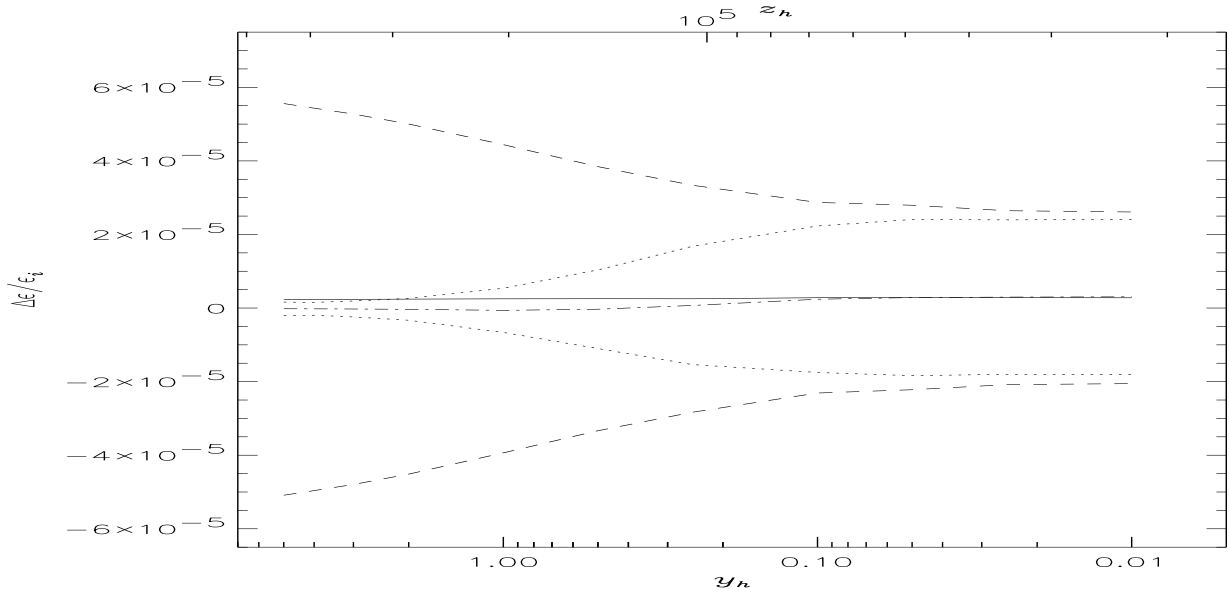,height=8cm,width=17cm}
\caption{Comparison between the constraints (at 95 per cent CL) 
on the energy exchanges
derived at different cosmic times from different data sets by considering
the case of a single dissipation process. 
The different lines refer to: 
FIRAS data calibrated according to Mather et al. 1999
(solid line: best fit; dashed lines: upper and lower limits); 
FIRAS data jointed to the 
data set D-BB (dot-dashes: best fit; dots: upper and lower limits).}
\end{figure*}

\subsubsection{Distorted spectra}\label{dimes_distorto}

The test
reported in the previous section suggests that even small distortions 
could be determined provided that the dissipation would have occurred at 
relatively early epochs,  $y_h \gsim 1$. 
Thus, we analyse the sensitivity of a DIMES-like experiment 
in the recovery of the amount of energy possibly injected 
in the radiation field and explore also the possibility to 
determine the dissipation process epoch. 

Firstly, we fit the data simulated as above under the hypothesis 
that the heating epoch is known; more explicitly, 
we fit the data with a theoretical spectrum distorted by a process 
occurring at the considered $y_h$ by allowing to optimize 
$\Delta\epsilon/\epsilon_i$ (and $T_0$) but by taking $y_h$ fixed.
In this way we can see how accurately
$\Delta\epsilon/\epsilon_i$ could be in principle recovered.

On the other hand, unless we want to use the CMB spectrum data to 
constrain theoretical models with a well defined dissipation epoch, 
we are typically interested to set constraints on 
the value of $\Delta\epsilon/\epsilon_i$ possibly injected at a given unknown 
epoch occurring within a relatively wide cosmic period;
in addition, many classes of physical processes 
in the plasma epoch involve time parameters and it is
important to understand how they can be 
possibly constrained by the comparison with CMB spectrum observations.

Thus, we focus on the cases of spectra 
distorted at high ($y_h=5$), medium ($y_h=1.5$) and at low 
($y_h\ll1$) redshifts by fitting the simulated data by relaxing the
a priori knowledge of the dissipation epoch. 
In this way we would be able to evaluate the possibility of 
determining also the epoch of the heating 
\footnote{Of course, any energy injection occurred at a certain 
$y_h > 5$ with a proper higher value of $\Delta \epsilon /\epsilon_i$ 
would give a distorted spectrum essentially indistinguishable
by that generated in the case of a dissipation at $y_h = 5$
with a lower value of $\Delta \epsilon /\epsilon_i$, see section~5.4.}
without a priori informations
by jointly evaluating the impact of the unknowledge of the dissipation
epoch on the recovery of injected energy.
We will test also the possibility of deriving at the same time
information on the baryon density.

\vskip 0.4cm 
\noindent
{\it 5.2.2.1 $\, \,$ Energy injections at FIRAS limits -- Dissipation epoch:
known}

\vskip 0.2cm 
\noindent
As a representative case we consider the simulated observation of a
spectrum distorted at different values of $y_h$ by an energy
injection with 
$\Delta\epsilon/\epsilon_i=2\times10^{-5}$, a 
value below, but not much, the FIRAS upper limit on 
$\Delta\epsilon/\epsilon_i$. 
These data are then compared with 
the theoretical CMB spectrum distorted at the same 
$y_h$ (assumed to be known)
by performing the fit only over 
$\Delta\epsilon/\epsilon_i$ and $T_0$:
this is appropriate to cases in which we have a 
quite well defined a priori information on the 
dissipation epoch but not on the amount of released energy.

We find that for high redshift processes,
$y_h \simeq 2-5$, $\Delta\epsilon/\epsilon_i$ is precisely determined.
For distortions at lower $z$, $y_h \lsim 1$, we obtain limits similar 
to those given from the currently available data, since the 
the fit result is mainly driven by FIRAS data, more sensitive
to these kinds of distortions, mainly
located at high frequencies.

\vskip 0.4cm 
\noindent
{\it 5.2.2.2 $\, \,$ Energy injections at FIRAS limits -- Dissipation epoch:
unknown}

\vskip 0.2cm 
\noindent
We relax here the assumption to know the dissipation epoch.
We consider firstly the case of the fit to data simulated assuming a
spectrum distorted at
$y_h=5$ with 
$\Delta\epsilon/\epsilon_i=2\times10^{-5}$
with CMB theoretical spectra distorted at different values of $y_h$.
The best-fit to these data assuming $y_h=5$ gives 
a very accurate recovery of the input value of
$\Delta\epsilon/\epsilon_i$
with a small quoted error (we find an 
associated statistical error of $\simeq 10$\% at 95 per cent CL).
The best-fit on 
$\Delta\epsilon/\epsilon_i$ assuming lower values of
$y_h$ is far from the input
value and the $\chi^2$ increases. 
Thus, we search for a favourite value
of $y_h$ by performing the fit over 
$\Delta\epsilon/\epsilon_i$, $T_0$, and $y_h$.
We obtain that the recovered best-fit value of $y_h$ is exactly 
the input one, 5.0, and lower limit on $y_h$ at 95 per cent CL is 2.4.
By searching also for a favourite
value of $\Ohat_b$ 
(set to 0.05 in the data simulation), we obtain
a 68 per cent CL range of $\simeq 0.01 - 0.096$.

We repeated the same analysis in the case of a spectrum
distorted at $y_h=1.5$ with 
$\Delta\epsilon/\epsilon_i=2\times10^{-5}$.
Again, the recovered value of $\Delta\epsilon/\epsilon_i$
is close to the input one for fits with $y_h \simeq 1.5$ 
(in this case we recover the input value 
of $\Delta\epsilon/\epsilon_i$
with an uncertainty of $\simeq 18$\% at 95 per cent CL)
and we are also
able to determine a significative range 
($y_h \gsim 1$ at 95 per cent CL) 
of favourite values
of $y_h$, although wider than in the previous case,
while $\Ohat_b$ is found to be in the range $\simeq 0.024 - 0.091$
at 68 per cent CL.

Similar results on $y_h$ and $\Ohat_b$ can not be
obtained in the case of fit to the data simulating 
the observation of a spectrum 
distorted at $y_h \sim 0.01$.
The fit result then 
is then compatible also with energy injections with smaller 
values of $\Delta\epsilon/\epsilon_i$ but at higher
$y_h$ and with a non distorted spectrum.
The $\chi^2$ value does not significantly 
change when $y_h$ varies. 
This is again the result of the main role
of FIRAS data for late dissipation processes.

\vskip 0.4cm 
\noindent
{\it 5.2.2.3 $\, \,$ Energy injections below FIRAS limits -- Dissipation
epoch: known}

\vskip 0.2cm 
\noindent
A DIMES-like experiment should be able to detect also small 
spectral distortions.
Let consider here the case of 
a spectrum distorted from an energy injection with 
$\Delta\epsilon/\epsilon_i=5\times10^{-6}$, 
about a factor 10 below the FIRAS limits at 95 per cent CL.

As shown in Fig.~3, if the dissipation epoch is known, we find
that for processes at early and intermediate epochs the best-fit result 
is very close to the input value of the simulated data,
although the limits on $\Delta\epsilon/\epsilon_i$ are not 
so stringent as in the case with a larger energy injection. 
For $y_h \gsim 1$, a spectral distortion would be firmly detected
at 95 per cent CL.

\begin{figure*}
\epsfig{figure=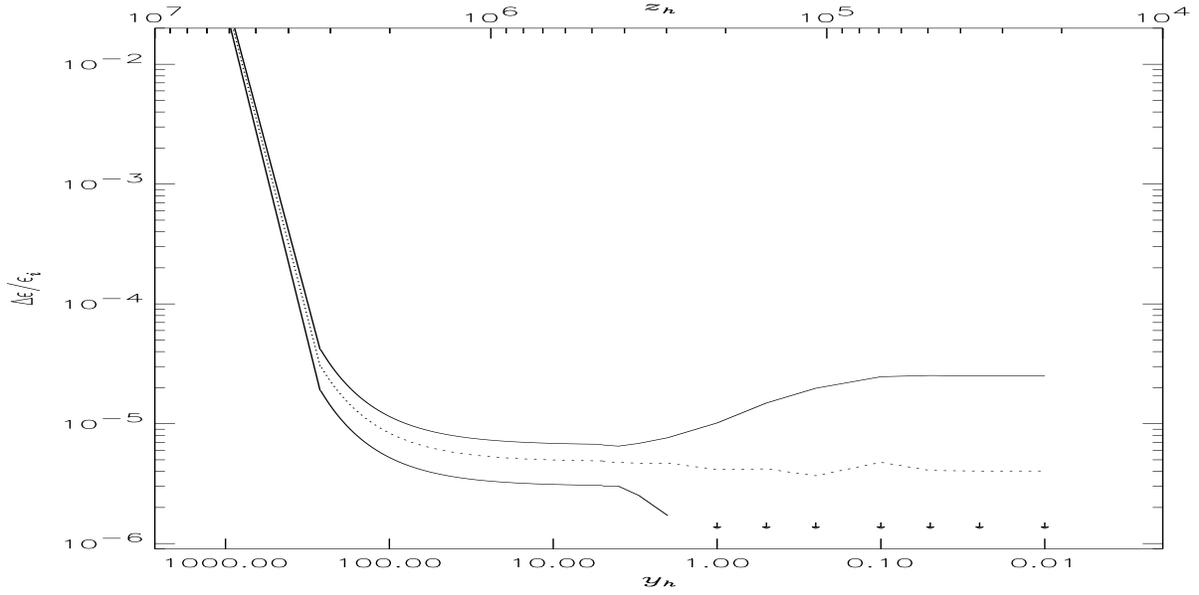,height=8cm,width=17cm}
\caption{Constraints on the energy exchanges
derived at different cosmic times by considering
the case of a single dissipation process
on the basis of the
FIRAS data calibrated according to Mather et al. 1999
and data simulated as in the case of an energy injection
with $\Delta\epsilon/\epsilon_i \ge 5\times10^{-6}$ 
and observed with a DIMES-like experiment.
The dissipation epoch is assumed to be known
(is the same in the generation of simulated data and in the fit).
The different lines refer to the best fit result 
(dots) and to the upper and lower limits at 95 per cent CL
(solid lines).
The arrows indicate that the sign of 
the lower limit changes at $y_h \simeq 1$, where lower and upper error bars 
result to be very similar.}
\end{figure*}

\vskip 0.4cm 
\noindent
{\it 5.2.2.4 $\, \,$ Energy injections below FIRAS limits -- Dissipation
epoch: unknown}

\vskip 0.2cm 
\noindent
We relax here again the assumption to know the dissipation epoch.
Our results are summarized in Fig.~4:
even for distortions well below the FIRAS limits
($\Delta\epsilon/\epsilon_i = 5\times10^{-6}$ is assumed here)
an experiment like DIMES would provide significative 
constraints both on the amount of dissipated energy and 
on the dissipation epoch 
in the case of and early processes. In this test 
the input dissipation epoch ($y_h =5$)
is again quite well recovered,
the $\chi^2$ increasing of 4 when $y_h$ becomes close to unity.
It would be also possible to provide an independent estimate
of the baryon density: we find $\Ohat_b \simeq 0.01 - 0.18$ 
at 68 per cent CL.

For a process occurring at intermediate epochs ($y_h=1.5$ 
in this specific test) we find that  
it is still possible to determine the amount of injected
energy, but in this case of distortions significantly smaller
than the FIRAS limits
the $\chi^2$ is no longer particularly sensitive to 
$y_h$ and the recovered range of dissipation epochs is wide.
Energy dissipations processes at intermediate epochs
may then result still compatible with these simulated data
and only energy injections at late epochs,
could be excluded 
(in this test we find that the $\chi^2$ increases of $\simeq 4$
for $y_h \simeq 0.1$).

As already found for larger distortions, 
significant information on 
late processes 
can not be obtained from accurate long wavelength data
because of the more relevant role of the FIRAS data. 

\begin{figure*}
\epsfig{figure=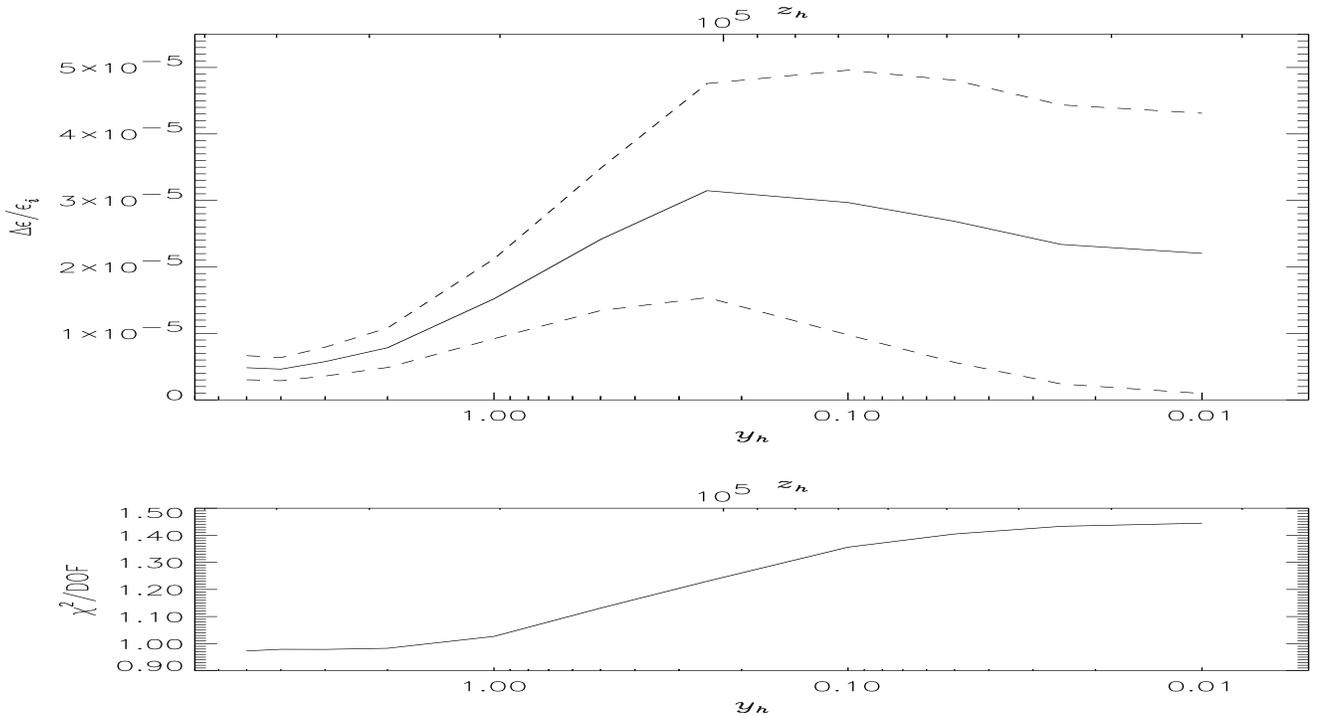,height=9cm,width=17cm}
\caption{Constraints on the energy exchanges
derived at different cosmic times by considering
the case of a single dissipation process
on the basis of the
FIRAS data calibrated according to Mather et al. 1999
and data simulated as in the case of an energy injection
with $\Delta\epsilon/\epsilon_i \ge 5\times10^{-6}$ 
occurring at $y_h=5$ 
and observed with a DIMES-like experiment (top panel).
The different lines refer to the best fit result 
(solid line) and to the upper and lower limits at 95 per cent CL
(dashes) by assuming a dissipation process occurring at 
a given value of $y_h$. 
Note that the $\chi^2/$d.o.f. increases
for decreasing $y_h$, i.e. when the dissipation epoch  
assumed in the fit is different from that adopted in the 
generation of the simulated data (bottom panel):
this implies that 
a DIMES-like experiment could provide interesting
constraints also on the epoch of a possible energy exchange.}
\end{figure*}

\vskip 0.4cm 
\noindent
{\it 5.2.2.5 $\, \,$ Very small energy injections}

\vskip 0.2cm 
\noindent
Finally, we consider the possibility to detect very small energy
injections, 
namely with $\Delta\epsilon/\epsilon_i = 2\times10^{-6}$. 

If the dissipation epoch is known, the result of the fit 
shows that is still possible
to determine a no null distortion provided that 
the dissipation process occurs at high redshifts. 

For energy injections at epochs close to $y_h=5$
the recovered value of 
$\Delta\epsilon/\epsilon_i$,
$\simeq 2.5\times10^{-6}$ in this test,  
is quite close to the input one 
and the associate statistical error gives
a $\Delta\epsilon/\epsilon_i$ range 
of $\simeq (1-4.5)\times10^{-6}$ at 95 per cent CL.
An indication of the dissipation epoch can be also
derived (in this case 
we find that the $\chi^2$ increases of $\simeq 4$ for $y_h \simeq 0.5$). 
Unfortunately, no significant informations on 
$\Ohat_b$ can be obtained 
with the considered sensitivity and frequency coverage
in the case of so small distortions.

For dissipations at $y_h \sim 1.5$ 
a significative distortion can be also determined, 
the recovered value of  $\Delta\epsilon/\epsilon_i$ ranging
between $\simeq 0$ and $\sim 8\times10^{-6}$ 
at 95 per cent CL for $y_h \sim 1-2$ where the 
$\chi^2/$d.o.f. is about  its minimum, 
but significative informations on the dissipation epoch 
can be no longer derived.

Finally, in the case of processes at late epochs, the 
fit results are compatible with an unperturbed spectrum, 
being these kind of distortions 
mainly detectable at FIRAS frequencies.

\subsection{Fits to simulated data: joint analysis of two dissipation
processes}

We discuss here the possibility to significantly improve
the constraints on (or to detect) energy exchanges also 
in the more general case of
a joint analysis of early/intermediate 
and late dissipation processes.

\subsubsection{Non distorted spectrum}

By exploiting the data set D-BB presented in Sect.~5.1
we consider the case of no significant deviations
from a Planckian spectrum.
Top panel of Fig.~5 shows 
the limits on the energy exchange 
as function of $y_h$
by allowing for a later dissipation process possibly occurred at
$y_h\ll1$; 
bottom panel of Fig.~5 shows
the constraints on the energy injected at low $z$ by allowing for 
a previous distortion occurred at any 
given $y_h$.
In Fig.~5 we report also the comparison with the results 
based only on the FIRAS data (as shown in Salvaterra \&
Burigana 2002, the current long wavelength measures do not
change significantly these results). 
The conclusion is impressive:
the constraints on $\Delta\epsilon/\epsilon_i$ 
for early and intermediate
dissipation processes could be improved by a factor
$\simeq 10 - 50$, depending on the considered dissipation epoch.
In addition, the constraints on the energy dissipation at 
late epochs can be also improved, by a factor of about two, because of
the reduction of the partial degeneracy 
introduced by the rough compensation 
(Salvaterra \& Burigana 2002) 
between the effect of
early and late energy exchanges on the CMB spectrum 
when no accurate measures are available at long wavelengths.

\begin{figure*}
\epsfig{figure=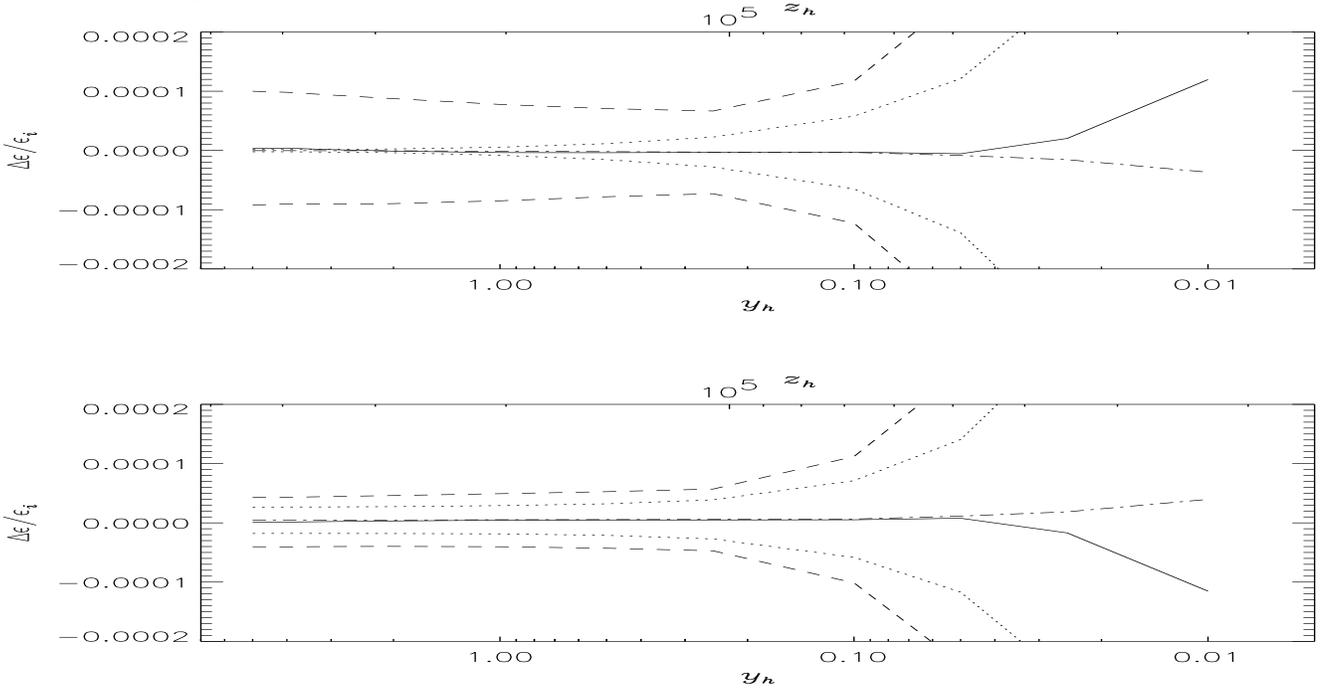,height=9cm,width=17cm}
\caption{Constraints at 95 per cent CL 
on the energy exchanges derived from current measures,
FIRAS and long wavelength data (in practice FIRAS data alone 
set the current constraints, 
see Salvaterra \& Burigana 2002;
solid lines: best fit; dashes:
upper and lower limits) and from FIRAS data jointed to 
the simulated data set D-BB (dash-dotted line: best fit; 
dots: upper and lower limits).
The top panel refers to the constraints on early/intermediate
energy exchanges occurring at the considered $y_h$
obtained by allowing for a possible 
late dissipation process.
The bottom panel refers to the constraints on late 
energy exchanges obtained by allowing for a possible 
dissipation process at an early/intermediate epoch
occurring at the considered $y_h$.
Note the large improvement (up to a factor $\simeq 50$) 
of the constraints on early/intermediate processes, 
because of the long wavelength coverage of DIMES-like
experiment, and that also the constraints on 
processes at late epochs could improve (by a factor $\simeq 2$).
See also the text.}
\end{figure*}

\subsubsection{Distorted spectra}

To complete the analysis of the impact of a possible DIMES-like experiment, 
we consider the simulated observation of a spectrum 
distorted by a first energy dissipation at $y_h=5$ with 
$\Delta\epsilon/\epsilon_i=5\times10^{-6}$ and a second one at $y_h\ll1$ with
$\Delta\epsilon/\epsilon_i=5\times10^{-6}$. 
We then compare these data with theoretical 
spectra distorted by a process at $y_h\ll1$ and another at any
given $y_h \lsim 5$ according to our grid of $y_h$
(see section~5.1). 

We find that a DIMES-like experiment would allow to firmly determine the
presence of the distortion at high $z$;
in particular, at $y_h=5$ the recovered $\Delta\epsilon/\epsilon_i$
is very close to the input one,
as already found for the tests described in sections~5.2.2.3 and
5.2.2.4.  
On the contrary, for the dissipation process at $y_h\ll1$ the fit result
is compatible with an unperturbed spectrum, since the FIRAS data dominate
the limits on the distortions at low redshifts. The limits on 
$\Delta\epsilon/\epsilon_i$ for processes at low redshifts are, however, 
again more stringent, by a factor 2, than those obtained with the
currently available data.

We find that the $\chi^2$ significantly 
increases by assuming in the fit an earlier process 
at decreasing $y_h$ (the $\chi^2$ increases of $\simeq 4$ 
for $y_h \simeq 1$): even in the case of 
a combination of an early and a late process
a DIMES-like experiment would be able to significantly 
constrain the epoch of the earlier energy exchange.

\subsection{Constraints on very high redshift processes}\label{dimes_evo}

We extend here at $z_h > z_1$ (i.e. $y_h > 5$)
the constraints on $\Delta\epsilon/\epsilon_i$ that 
would be possible to derive 
at $z_h = z_1$ ($y_h = 5$) with a DIMES-like experiment.
We remember that at $z>z_1$ the Compton scattering 
is able to restore, after an energy 
injection, the kinetic equilibrium between matter and radiation, yielding a 
Bose-Einstein (BE) spectrum, and the combined effect of Compton scattering 
and photon production processes tends to reduce the magnitude 
of spectral distortions, possibly leading to a blackbody spectrum. 

We firstly consider here the case of the simulated observation of a not distorted spectrum, 
that represents a good test of the possible improvements of an instrument like 
DIMES, because the limits on $\Delta\epsilon/\epsilon_i$ at relevant redshifts 
can be directly compared to those obtained with FIRAS data alone.
For simplicity, we consider the case of a single energy injection
possibly occurred in the cosmic thermal history.
The comparison is shown in Fig.~6.
As evident, the constraints on $\Delta\epsilon/\epsilon_i$ can be
improved by a factor $\sim 10 - 50$ for processes possibly occurred
in a wide range of cosmic epochs, corresponding to about a decade 
in redshift at $z$ about $10^6$.
Of course, large energy injections are still possible at very early epochs 
close to the thermalization redshift, when primordial nucleosynthesis 
set the ultimately constraints on energy injections in the cosmic radiation field.
For late dissipations, FIRAS data mainly constrain $\Delta\epsilon/\epsilon_i$.

As a further example, we consider 
the constraints on the energy injections at $z_h \ge z_1$ 
($y_h \ge 5$)
in the case of a fit with a single energy injection to 
simulated observations of a spectrum distorted at $y_h = 5$
with $5\times10^{-6}$. 
As shown by the high redshift tails of the curves of Fig.~3, in this case
the constraints on the thermal history of the Universe would be
completely different from those 
derived in the case in which distortions are not detected (Fig.~6).
A firm detection of early energy injections would be clearly possible with
the considered experimental performances and the constraints on 
the energy possibly injected at $z>z_1$ could be directly derived 
from such kind of future CMB spectrum data.
In addition, Fig.~6 shows that, contrariously to the case of the 
current observational status, 
the constraints on early energy exchanges
based on future high accuracy long wavelength measures
are no longer appreciably relaxed by assuming that a late process could be also 
occurred.

\begin{figure*}
\epsfig{figure=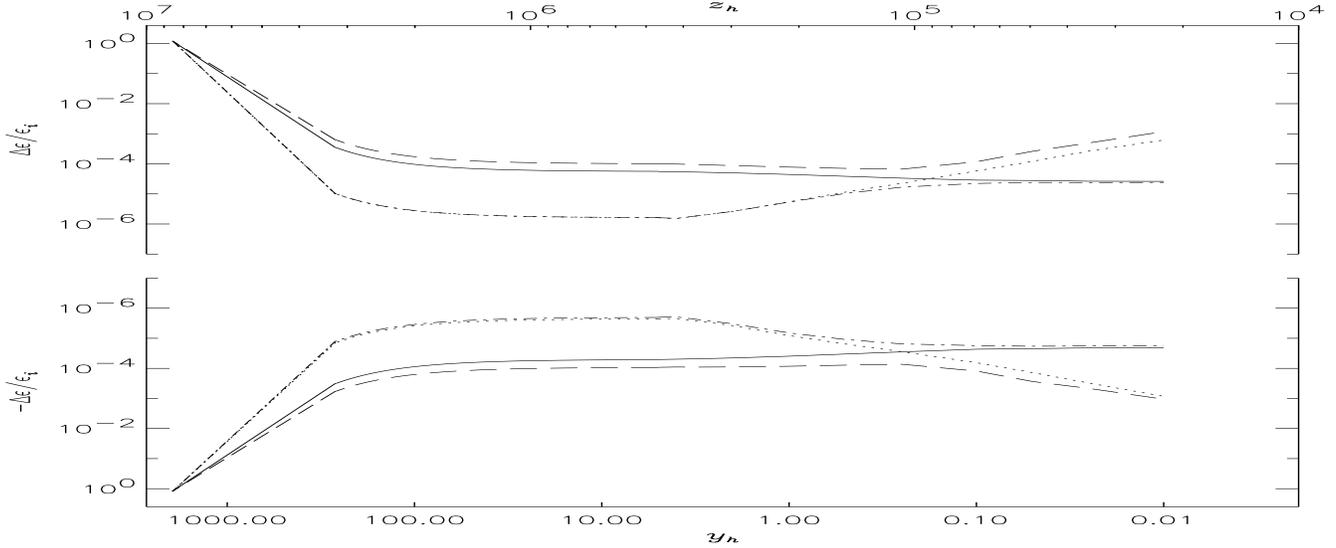,height=8cm,width=17cm}
\caption{Constraints on the energy exchanges derived from current measures,
FIRAS and long wavelength data (solid lines and long dashes; 
in practice FIRAS data alone set the current constraints, 
see Salvaterra \& Burigana 2002) and from FIRAS data jointed to 
the simulated data set D-BB (dash-dotted lines and dots).
In both cases, the pairs of lines with the more stringent constraints
refer to the case in which a single energy exchange is considered,
while the others refer to the case of the joint analysis
of early/intermediate and late processes.}
\end{figure*}

\section{Energy injections above ``standard'' FIRAS limits and 
check of FIRAS calibration}


In the previous sections, we considered simulated data 
with a Planckian spectrum or with distortions compatible
with the limits derived from FIRAS data calibrated
according to Mather et al. 1999.
On the other hand, 
a recent analysis of FIRAS calibration 
(referred here as ``revised'')
by Battistelli et~al. 2000
suggests a frequency dependence 
of the FIRAS main calibrator emissivity.
The FIRAS data recalibrated according to their ``favourite'' 
calibration emissivity law (R-FIRAS data in what follows)
indicate the existence of deviations from a Planckian shape 
or at least a significant relaxation of the constraints 
on them. 
Salvaterra \& Burigana 2002 discussed the main implications
of this analysis. Although it seems quite difficult to fully
explain from a physical point of view the R-FIRAS data,
two classes of phenomenological models 
may fit them: in the first one the main contribution
derives from an intrinsic CMB spectral distortion  
with $\Delta\epsilon/\epsilon_i \simeq$~few~$\times 10^{-4}$
occurring at early/intermediate epochs; 
the second one involves a millimetric component possibly due to cold 
dust emission, 
described by a modified blackbody spectrum, added to a CMB blackbody 
spectrum at a temperature of $\simeq 4$~mK below 
FIRAS temperature scale of 2.725~K. 
Fig.~7 shows as these models,
very similar at millimeter wavelengths, predict 
significant differences at centimeter and decimeter wavelengths.

In this section we carefully discuss the capabilities 
of forthcoming and future CMB spectrum measures
at long wavelengths to discriminate between these two different FIRAS 
calibrations and, in the case of the 
calibration by Battistelli et~al. 2000, 
to distinguish between the two above scenarios.


%
\begin{figure*}
\epsfig{figure=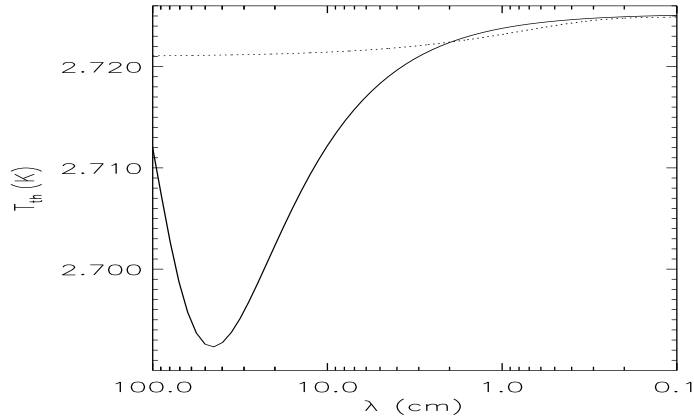,height=6cm,width=10cm}
\caption{The CMB spectrum as would observed today (here expressed 
in terms of equivalent thermodynamic
temperature) in the case an early distortion occurring
at $y_h=5$ with $\Delta\epsilon/\epsilon_i = 2\times 10^{-4}$ 
(solid line) and in the case of a sum of a pure blackbody
spectrum plus a modified blackbody spectrum, representative 
of typical dust emission, with the parameters of Table~14
of Salvaterra \& Burigana 2002 (dots), two models that 
fit the FIRAS data as recalibrated by Battistelli et al. 2000.
Note that these two models, very similar at 
$\lambda \lsim 1$~cm, differ of 
$\simeq$~some~mK~$-$~30~mK at 
$\lambda \sim$~some~cm~$-$~some~dm.}
\end{figure*}

\subsection{Ruling out the ``revised'' FIRAS calibration}

In the case in which the FIRAS calibration by Mather et al. 1999
is substantially correct, and therefore that revised by Battistelli et al.
2000 is wrong,
the CMB spectrum is expected to show an essentially Planckian shape also 
at $\lambda \gsim 1$~cm. To test the capabilities of forthcoming 
and future CMB spectrum experiments to rule out the ``revised'' calibration
we fit simulated observations of a Planckian spectrum
at $\lambda \gsim 1$~cm generated as described in section~3.1 
added to the R-FIRAS data set in terms of a single energy exchange process
and of a combination of two processes at different cosmic times.
We consider long wavelength data simulated assuming different values
of $T_0$, at steps of 0.1~mK within the FIRAS temperature 
scale range, in order to reach the best compromise between them and the
non-flat R-FIRAS temperature data. 
The first five rows of Table~1 summarize our results 
in the case of a DIMES-like experiment assuming 
$T_0 \simeq 2.7248$~K, the case in which we find the best agreement.
For comparison, we report 
in the last three rows of Table~1 
the results obtained by considering
the R-FIRAS data alone (Salvaterra \& Burigana 2002).
As evident from the significant increase of the $\chi^2/$d.o.f., 
long wavelength measures indicating a CMB Planckian shape
are not compatible with the FIRAS data calibrated according to
Battistelli et al. 2000.
Similar analyses carried out in the case of simulated long wavelength 
Planckian data with the sensitivity of forthcoming, or improved, ground and 
balloon experiments (data sets G\&B1-BB and G\&B2-BB) 
show values of $\chi^2/$d.o.f. close to $1.1-1.3$ (similar to those
obtained considering R-FIRAS data alone),
when the epoch (or the epochs) and the energy exchange (exchanges)
of the dissipation process (processes)
is (are) properly chosen. 

We then conclude that a CMB spectrum experiment 
at long wavelengths designed to rule out the FIRAS calibration
as revised by Battistelli et al. 2002 should have a sensitivity
comparable to that of a DIMES-like experiment.

\begin{table}
\begin{center}
\begin{tabular}{llll}
\hline
\hline
\multicolumn{3}{c}{$(\Delta\epsilon/\epsilon_i)/10^{-6}$} & $\chi^2/$d.o.f.  \\
\cline{1-3}
$y_h=5$ & $y_h=1.5$ & $y_h\ll 1$ & \\ 
\hline
$1.75\pm1.81$ & & & 2.401 \\
& $4.66\pm3.81$ & & 2.357 \\
& & $62.37\pm21.02$ & 1.784 \\
$-0.06\pm1.93$ & & $62.61\pm22.31$ & 1.819 \\
& $1.22\pm4.17$ & $62.10\pm22.96$ & 1.819 \\
\hline
\hline
$201.5\pm53.2$ & & & 1.248 \\
& & $60.7\pm23.0$ & 1.906 \\
$278.3\pm96.6$ & & $-39.9\pm41.8$ & 1.194 \\
\hline
\end{tabular}
\end{center}
\caption{Results of the fit to the R-FIRAS data combined 
(first five rows) or not (last three rows) with 
long wavelength data with a sensitivity of a DIMES-like
experiment and simulated according to a Planckian shape
at $T_0=2.7248$~K (first 5 rows) in terms of a single or 
two dissipation processes at different epochs 
(fits to two or three parameters: $T_0$ and one or two values of 
$\Delta\epsilon/\epsilon_i$). See also the text.}
\label{tab:3}
\end{table}

\subsection{Ruling out the ``standard'' FIRAS results}

In the case in which the FIRAS calibration revised by Battistelli
et al. 2000 is substantially correct, and therefore that by Mather et al.
1999 is wrong,
the CMB spectrum is expected to show deviations from the Planckian shape 
at $\lambda \gsim 1$~cm, as shown in Fig.~7. 
To test the capabilities of forthcoming 
and future CMB spectrum experiments to rule out the ``standard'' FIRAS
calibration  we consider simulated observations of spectra 
distorted by an early energy injection or by a significant
cold dust contribution, able to account for the FIRAS calibration by 
Battistelli et al. 2000. Of course, this analysis is useful also
to test the capability of future experiments to cross-check the
``standard'' FIRAS results, independently of the FIRAS
calibration revision by Battistelli et al. 2000.

Firstly, we consider the case of long wavelength data simulated 
according to an early, BE-like distortion occurring at $y_h=5$ 
with $\Delta\epsilon/\epsilon_i = 2 \times 10^{-4}$.
We exploit three sets of data, the first two with the sensitivities described 
in section~4.1 (G\&B1-BE and G\&B2-BE) and the third
with the sensitivity of a DIMES-like experiment (D-BE).
We consider these data separately and combined with the FIRAS data
calibrated according to Mather et al. 1999 and to Battistelli
et al. 2000. The result is reported in Table~2.

As evident, measures with sensitivities similar to those
of the data set G\&B1-BE (see rows $1-3$ of Table~2) are not able to recognize
the inconsistency between the Planckian shape of the ``standard'' FIRAS
data and the long wavelength data, assumed distorted. Also,
they can not clearly support the ``revised'' FIRAS calibration.

On the contrary, with the very high sensitivity of a DIMES-like experiment 
(see rows $7-9$ of Table~2) 
it could possible to accurately measure the amount 
of energy dissipated at early
cosmic times and accurately cross-check the FIRAS calibration.

Finally, it may be interesting to note that an improvement 
by a factor $\sim 10$ of the sensitivity 
with respect to that of forthcoming ground experiments
(as in the case of the data set G\&B2-BE; see rows $4-6$ of Table~2) 
could provide a significant, although not fully exhaustive,
independent test of the FIRAS calibration.

\begin{table}
\begin{center}
\begin{tabular}{lll}
\hline
\hline
Data set & $(\Delta\epsilon/\epsilon_i)/10^{-4}$ & $\chi^2/$d.o.f.  \\
\hline
G\&B1-BE & $1.64\pm9.96$ & 0.693 \\
G\&B1-BE + FIRAS & $0.04\pm0.53$ & 0.917 \\
G\&B1-BE + R-FIRAS & $2.01\pm0.53$ & 1.107 \\
G\&B2-BE & $1.96\pm0.99$ & 0.693 \\
G\&B2-BE + FIRAS & $0.83\pm0.41$ & 1.301 \\
G\&B2-BE + R-FIRAS & $1.98\pm0.41$ & 1.108 \\
D-BE & $1.99\pm0.03$ & 1.056 \\
$^{(a)}$~D-BE + FIRAS & $1.99\pm0.02$ & 2.041 \\
D-BE + R-FIRAS & $1.98\pm0.02$ & 1.215 \\
\hline
\end{tabular}
\end{center}
\caption{Results of fits to long wavelength data 
at different sensitivities
simulating the observation of an early energy injection occurring at $y_h=5$ 
with $\Delta\epsilon/\epsilon_i=2\times10^{-4}$ combined 
to the FIRAS data (rows 2, 5, 8), the R-FIRAS data (rows 3, 6, 9)
or alone (rows 1, 4, 7).
Note that the $\chi^2/$d.o.f. 
of a BE-like spectrum with $\Delta\epsilon/\epsilon_i=2\times10^{-4}$
compared to the ``standard'' FIRAS data alone is $\simeq 2.177$
(fits to two parameters: $T_0$ and $\Delta\epsilon/\epsilon_i$;
errors at 95 per cent CL). 
See also the text.
$^{(a)}$~As generalization of this test, 
we verified that 
for a wide range of $y_h$ (between $y_h=5$ and $y_h \ll 1$)
the FIRAS data are not compatible 
with long wavelength data simulated assuming distortions
with $\Delta\epsilon/\epsilon_i = 2 \times 10^{-4}$
and observed with the sensitivity of a DIMES-like experiment, 
since the large $\chi^2/$d.o.f. values found even 
in the most favourite cases in which the dissipation epoch is assumed 
known.}
\label{tab:distorto}
\end{table}

As a second test, we consider the case of long wavelength data simulated 
according to a Planckian shape plus the contribution from a 
modified blackbody emission described by the parameters 
of Table~14 of Salvaterra \& Burigana 2002 combined to ``standard'' FIRAS 
data at $\lambda \lsim 1$~cm.
We exploit three sets of data, the first two with the sensitivities described 
in section~4.1 (G\&B1-BB/D and G\&B2-BB/D) and the third
with the sensitivity of a DIMES-like experiment (D-BB/D), and
fit them in terms of a pure Planckian spectrum (fits on $T_0$). 
The result is reported in Table~3.

As evident, with the very high sensitivity of a DIMES-like experiment 
it could be possible to probe the disagreement between the long wavelength
region of the CMB spectrum and the subcentimetric FIRAS data when calibrated 
according to Mather et al. 1999. The next ground and balloon experiment
do not have enough sensitivity, while a significant, 
although not fully exhaustive,
independent test could be obtained by improving their 
sensitivities by a factor $\sim 10$.

\begin{table}
\begin{center}
\begin{tabular}{ll}
\hline
\hline
Data set & $\chi^2/$d.o.f. \\
\hline
G\&B1-BB/D + FIRAS & 0.981 \\
G\&B2-BB/D + FIRAS & 1.325 \\
D-BB/D + FIRAS & 91.10 \\
\hline
\end{tabular}
\end{center}
\caption{Results of fits to long wavelength data at different sensitivities 
simulating the observation of a Planckian spectrum plus a contribution 
from a modified blackbody spectrum described by the parameters of Table~14 
of Salvaterra \& Burigana 2002 combined to FIRAS data which show a Planckian 
shape (fits on $T_0$ assuming a pure blackbody spectrum). 
Note how the different experimental sensitivity 
allows or not to test the disagreement between the spectral regions at 
$\lambda$ longer or shorter than $1$~cm. See also the text.}
\label{tab:distorto}
\end{table}

\subsection{Discriminating from different scenarios for the 
``revised'' FIRAS calibration}

As discussed in sections~6.1 and 6.2, accurate long wavelength measures
could provide independent cross-checks of FIRAS calibration, a 
DIMES-like experiment being in principle able to firmly rule out 
the ``revised'' FIRAS calibration as well as the ``standard'' one. 
In the latter case, it is interesting to understand the experimental
requirements to discriminate between different scenarios to account
for the ``revised'' FIRAS calibration.

To address this point we consider two representative tests:
in the first case we assume an energy injection 
occurred at early times and try to explain the resulting
spectrum in terms of a CMB Planckian spectrum plus a relevant
millimetric astrophysical foreground, in the second one we assume 
a CMB Planckian spectrum plus a relevant millimetric astrophysical foreground
and try to explain the resulting
spectrum in terms of an early energy injection. 

The results of the first test are summarized in Table~4 for our
three cases with different long wavelength sensitivities.
In this case we re-add (in terms of antenna temperature) 
the isotropic component (F96) found by Fixsen et al. 1996 
(see also section~7 of Burigana \& Salvaterra 2002) 
to the monopole data 
and consider the simulated long wavelength data 
combined or not to the R-FIRAS data.

As evident from the very large values the $\chi^2/$d.o.f.,
a DIMES-like experiment could firmly rule out the model involving a relevant
millimetric component described by a modified blackbody
to account for the ``revised'' FIRAS calibration,
both considered alone and in combination with 
the R-FIRAS data, so supporting the possibility of a relevant
early energy injection to account for the ``revised'' FIRAS calibration.
Again, next ground and balloon experiments do not have enough
sensitivity, while an improvement by a factor $\sim 10$ 
of their sensitivity may point out the difficulty of a cold dust
model in jointly explaining long and short wavelength data.

\begin{table}
\begin{center}
\begin{tabular}{ll}
\hline
\hline
Data set & $\chi^2/$d.o.f. \\
\hline
G\&B1-BE/F96 & 0.636 \\
G\&B1-BE/F96 + R-FIRAS & 0.885 \\
G\&B2-BE/F96 & 1.705 \\
G\&B2-BE/F96 + R-FIRAS & 1.107 \\ 
D-BE/F96 &  3231 \\
D-BE/F96 + R-FIRAS & 499.3 \\
\hline
\end{tabular}
\end{center}
\caption{Results of fits to long wavelength data at different sensitivities 
simulating the observation of an early energy injection occurred at $y_h=5$ with 
$\Delta\epsilon/\epsilon_i=2\times10^{-4}$ combined or not to R-FIRAS 
data (the isotropic foreground F96 is re-added). The fit model includes
the two component astrophysical monopole of Table~14 of Salvaterra \& Burigana 2002
leaving only $T_0$ as a free parameter.
A DIMES-like experiment could clearly rule out the model involving a relevant
millimetric component described by a modified blackbody
to account for the ``revised'' FIRAS calibration,
while forthcoming ground experiments do not have enough sensitivity.
See also the text.}
\label{tab:distorto}
\end{table}

The results of the second test are summarized in Table~5, again for 
our three cases with different long wavelength sensitivities.

As evident from the very large $\chi^2/$d.o.f., 
a DIMES-like experiment could firmly rule out the model involving an 
early energy injection even allowing for a proper adjustment 
of the value of $\Delta\epsilon/\epsilon_i$, so 
supporting the model involving a significant millimetric
contribution to account for the ``revised'' 
FIRAS calibration, while much less accurate experiments 
do not provide robust answers.

\begin{table}
\begin{center}
\begin{tabular}{lll}
\hline
\hline
Data set & $(\Delta\epsilon/\epsilon_i)/10^{-4}$ & $\chi^2/$d.o.f. \\
\hline
G\&B1-BB/D &$3.41\pm10.18$ & 1.042 \\
G\&B2-BB/D &$0.57\pm0.99$ & 1.108 \\
D-BB/D & $0.28\pm0.03$ & 63.09\\
\hline
\end{tabular}
\end{center}
\caption{Results of fits to long wavelength data at different sensitivities 
simulating the observation of a Planckian spectrum plus a modified blackbody
according to Table~14 of Salvaterra \& Burigana 2002 in terms 
of an early spectral distortion occurring at $y_h=5$ (fits to
$T_0$ and $\Delta\epsilon/\epsilon_i$; errors at 95 per cent CL).
A DIMES-like experiment could clearly rule out the model 
involving an early distortion to account for the ``revised'' 
FIRAS calibration. See also the text.}
\label{tab:distorto}
\end{table}

\section{Free-free distortions}\label{free-free}

Long wavelength measures are particularly sensitive to the 
free-free distortion due to its well known dependence 
on the frequency

\begin{equation}
{T_{th}-T_0\phi_i \over T_0} \simeq {y_B \over x^2} \, ,
\end{equation}

\noindent
where $\phi_i$ is the ratio between the electron and radiation 
temperature before the beginning of the dissipation process,
$y_B$ is the so called free-free distortion 
parameter (Burigana et al. 1995), and $T_{th}$ is the (frequency dependent)
equivalent thermodynamic temperature.
The current constraints on the $\vert y_B \vert$ 
depend on the particular set of measures included 
in the analysis and typically range between
$\sim$~few~$\times 10^{-5}$ and $\sim 10^{-4}$ 
at 95 per cent CL (Salvaterra \& Burigana 2002).

In Table~6 we report the best fit values 
of $y_B$ with the errors at 95 per cent CL 
and the values of $\chi^2/$d.o.f. for different data
sets simulated as described in section~3.1 
assuming, within the quoted errors, a Planckian spectrum 
at a temperature of 2.725~K,
combined or not with FIRAS data calibrated according
to Mather et al. 1999.

As evident, future long wavelength measures  
can significantly improve the current observational status:
constraints on (or detection of) $y_B$ at the (accuracy) level 
of $\sim 1-2 \times 10^{-5}$
can be reached by forthcoming experiments,
while improving the sensitivity up to that of a DIMES-like
experiment will allow to measure $y_b$ with an accuracy
up to about $10^{-7}$ (errors at 95 per cent CL). 

\begin{table}
\begin{center}
\begin{tabular}{lll}
\hline
\hline
Data set & $y_B$ & $\chi^2/$d.o.f. \\
\hline

G\&B1-BB & $(0.48\pm1.38) \times 10^{-5}$ & 1.027\\
G\&B1-BB + FIRAS & $(0.46\pm1.35) \times 10^{-5}$ & 0.977 \\
G\&B2-BB & $(0.48\pm1.37) \times 10^{-6}$ & 1.027 \\
G\&B2-BB + FIRAS & $(0.42\pm1.36) \times 10^{-6}$ & 0.979\\
D-BB & $(0.05\pm1.05) \times 10^{-7}$ & 1.173 \\
D-BB + FIRAS &  $(-0.23\pm0.88) \times 10^{-7}$ & 0.976 \\
\hline
\end{tabular}
\end{center}
\caption{Sensitivity of future long wavelength experiments 
to the measure of the free-free distortion parameter 
$y_B$ (errors at 95 per cent CL). See also the text.} 
\label{tab:yB}
\end{table}

\section{Experimental requirements for a baryon density evaluation}\label{dimes_omegab}

For distortions at relatively high redshifts ($y_h \gsim 1$), 
the value of $\Ohat_b$ can be simply determined by
the knowledge of the frequency position of the minimum of 
the CMB absolute temperature:

\begin{equation}\label{eq:37b}
\lambda_{m, BE} \simeq 41.56 \left(\frac{\Ohat_b}{0.05}\right)^{-2/3}\,\hbox{cm} \, .
\end{equation}

This opportunity is very powerful in principle, since the dependence of 
$\lambda_{m, BE}$ on $\Ohat_b$ is determined only by the well known 
physics of the radiation processes in an expanding Universe during the
radiation dominated era.

For dissipations at $y_h \gsim 5$, 
the amplitude of this temperature decrement is

\begin{equation}\label{eq:dTb}
\Delta T_m \simeq 1.17\times10^{-3}\left(\frac{\mu_0}{10^{-5}}\right)
\left(\frac{T_0}{2.725\,\mbox{K}}\right)
\left(\frac{\Ohat_b}{0.05}\right)^{-2/3}\,\mbox{K} \, ,
\end{equation}

\noindent
where $\mu_0$ is the chemical potential at the redshift $z_1$ corresponding 
to $y_h=5$.  

Eq.~(7) gives the range of wavelengths to observe 
for a firm evaluation of $\Ohat_b$. As example, 
for $\Ohat_b=0.05$ we need to accurately measure the CMB
absolute temperature up to wavelengths of about 50~cm, clearly out from
the DIMES range.
Ground and ballon experiments are currently planned to reach these wavelengths.
Moreover, the amplitude of the
maximum dip of the brightness temperature 
for the energy dissipations at $y_h \gsim 5$, see Eq.~(8), 
turns to be at the mK level
for $\Ohat_b \sim 0.05$ and distortions within the FIRAS limits;
Burigana et al.~1991a shown that it is 
about 3 times smaller for energy injections at $y_h \simeq 1$. 
Experiments designed to estimate $\Ohat_b$ through the 
measure of $\lambda_{m, BE}$ should then have a sensitivity level 
of $\sim 1$~mK or better.

For sake of illustration, we consider the simulated observation 
of a spectrum distorted at $y_h=5$ 
with $\Delta\epsilon/\epsilon_i=2\times10^{-5}$
(or $\Delta\epsilon/\epsilon_i=2\times10^{-4}$, as  suggested 
by the ``revised'' FIRAS data)
in a $\Ohat_b=0.05$ Universe
through a very precise  experiment extended up to $\lambda \sim 70$~cm.
More precisely, we consider the DIMES channels 
combined to measures at 
73.5, 49.1, 36.6, 21.3, 12 and 6.3~cm 
as those proposed for the space experiment LOBO dedicated to measure 
the CMB spectrum at very low frequencies ($0.408-4.75$~GHz; see
Sironi et al. 1995, Pagana \& Villa 1996), but
we assume a much better sensitivity, $\simeq 0.1$~mK, 
comparable to that 
of the DIMES-like experiment, (or $\simeq 1$~mK). 
Again, we generate the simulated data
as described in section~2.3.
The fit to these simulated data by assuming to know the dissipation epoch
shows that it would be possible to accurately determine 
both the amount of injected energy and the baryon density. We recover
$\Delta\epsilon/\epsilon_i=(2.11\pm0.14)\times10^{-5}$  
(or $\Delta\epsilon/\epsilon_i=(2.11\pm0.14)\times10^{-4}$) and 
$\Ohat_b=0.053\pm0.004$ (errors at 95 per cent CL).

Unfortunately, experiments at decimeter wavelengths with a 
sensitivity of $\sim 1$~mK or better,
although very informative in principle,
seem to be very far from current possibilities.

\section{Conclusions}

We have studied the implications of possible future observations of the CMB 
absolute temperature at $\lambda\gsim 1$~cm, where 
both ground, balloon and 
space experiments are currently under study to complement 
the accurate FIRAS data at $\lambda \lsim 1$~cm.

Our analysis shows that future measures from 
ground and balloon will not be able to significantly improve 
the constraints on energy exchanges in the primeval plasma already provided
by the FIRAS data.
Even observations with a sensitivity better by a factor 10 with respect to
the realistic performances of the next experiments at different 
centimeter and decimeter wavelengths
can not significantly improve this conclusion.

Thus, we have studied the impact of very high quality data, such those that
could be in principle reached with a space experiment.
For this analysis, we referred to the DIMES experiment (Kogut 1996),
submitted to the NASA in 1995, planned to measure the CMB absolute 
temperature at $\lambda \sim 0.5 - 15$~cm with a sensitivity of
$\simeq 0.1$~mK,
close to that of FIRAS.

We have demonstrated that these
data would represent a substantial improvement for our knowledge 
of energy dissipation processes at intermediate and high 
redshifts ($y_h \gsim 1$). 

Dissipation processes at $y_h \simeq 5$ 
could be accurately constrained and possibly firmly detected even for very small 
amounts of the injected energy 
($\Delta\epsilon/\epsilon_i \sim 2\times10^{-6}$). For these early
dissipation processes it would be possible to estimate also the energy injection epoch.
Distortions at intermediate redshifts ($y_h \sim 1.5$) could be 
also firmly detected, although 
in this case interesting information on the heating epoch can be derived
only for energy injections, $\Delta\epsilon/\epsilon_i$, 
larger than about $10^{-5}$.

On the contrary, by considering the case of a single energy exchange
in the thermal history of the Universe,
for late processes ($y_h \lsim 0.1$) a such kind of experiment can not 
substantially improve the limits based on the FIRAS data 
at $\lambda \lsim 1$~cm,  
which would still set the constraints on $\Delta\epsilon/\epsilon_i$ 
at late epochs.

By the jointed analysis of two dissipation processes occurring at different epochs,
we demonstrated that 
the sensitivity and frequency coverage of a DIMES-like experiment would allow 
to accurately recover the amount of energy 
exchanged in the primeval plasma
at early and intermediate redshifts, and possibly the corresponding epoch, 
even in presence of a possible late distortion.
Even in this case, 
the constraints on $\Delta\epsilon/\epsilon_i$ can be
improved by a factor $\simeq 10-50$ for processes possibly occurred
in a wide range of cosmic epochs, corresponding to about one--two decades 
in redshift at $z$ about $10^6$, while 
the constraints on the 
energy possibly dissipated at late epochs 
can be also improved by a factor $\simeq 2$,
because the rough compensation between the distortion effects
at millimetric wavelengths from an early and a late process with opposite signs
becomes much less relevant in presence of 
very accurate long wavelength data.

In addition, accurate long wavelength measures can provide an independent
cross-check of the FIRAS calibration: a DIMES-like experiment could accurately
distinguish between the FIRAS calibrations by Mather et al. 1999 
and by Battistelli et al. 2000 and in this second case could discriminate 
between different scenarios to account for it. Interesting, although not 
fully exhaustive, indications on this aspect could be also obtained
by improving the sensitivity of the next ground and balloon experiments
by a factor $\sim 10$. 
Further, we have shown that a possible accurate observation 
of spectral distortions at $\lambda > 1$~cm 
compatible with relatively large energy injections, 
compared to the ``standard'' FIRAS limits,
can not be consistently reconciled with the FIRAS data, at least for the class
of distortion considered here. In this observational scenario, ``exotic'' models
for spectral distortions should be carefully considered.

We have shown that future long wavelength measures  
can significantly improve the current observational status
of the free-free distortion:
constraints on (or detection of) $y_B$ at the (accuracy) level 
of $\sim 1-2 \times 10^{-5}$
can be reached by forthcoming experiments,
while improving the sensitivity up to that of a DIMES-like
experiment will allow to measure $y_b$ with an accuracy
up to about $10^{-7}$ (errors at 95 per cent CL). 

Of course, not only a very good sensitivity, but also an extreme 
control of the all systematical effects and, in particular, of the frequency calibration
is crucial to reach these goals.

Finally, 
a DIMES-like experiment will be able to provide indicative 
independent estimates of the baryon density: the 
product $\Omega_b H_0^2$ can be recovered 
within a factor $\sim 2 - 5$ even in the case of (very small) early
distortions with $\Delta\epsilon/\epsilon_i \sim (5 - 2) \times 10^{-6}$.
On the other hand, for $\Omega_b (H_0/50)^2 \lsim 0.2$,
an independent baryon density 
determination with an accuracy
at $\sim$ per cent level, comparable to that achievable
with CMB anisotropy experiments, 
would require an accuracy of $\sim 1$~mK or better in the measure 
of possible early distortions but 
up to a wavelength from $\sim$~few~$\times$~dm to $\sim 7$~dm, according
to the baryon density value.

\section*{Acknowledgements}

It is a pleasure to thank M.~Bersanelli, N.~Mandolesi, 
C.~Macculi, G.~Palumbo, and G.~Sironi  
for useful discussions on CMB spectrum observations.
C.B. warmly thank L.~Danese and G.~De~Zotti
for numberless conversations on theoretical aspects
of CMB spectral distortions. 


\begin{thebibliography}{}

\bibitem[]{} Battistelli E.S., Fulcoli V., Macculi C. 2000, New Astronomy, 5, 77
\bibitem[]{} Burigana C., Danese L., De Zotti G. 1991a, A\&A, 246, 59
\bibitem[]{} Burigana C., De Zotti G., Danese L. 1991b, ApJ, 379, 1
\bibitem[]{} Burigana C., De Zotti G., Danese L. 1995, A\&A, 303, 323
\bibitem[]{} Burigana C. \& Salvaterra R. 2000, Int. Rep. ITeSRE/CNR 291/2000, August
\bibitem[]{} Danese L. \& Burigana C. 1993, in: ``Present and Future
of the Cosmic Microwave Background'', Lecture in Physics, Vol. 429, eds.
J.L. Sanz, E. Martinez-Gonzales, L. Cayon, Springer Verlag, Heidelberg (FRG), p. 28
\bibitem[]{} Danese L. \& De Zotti G. 1977, Riv. Nuovo Cimento, 7, 277
\bibitem[]{} Fixsen D.J. et al. 1994, ApJ, 420, 457
\bibitem[]{} Fixsen D.J. et al. 1996, ApJ, 473, 576
\bibitem[]{} Kogut A. 1996, ``Diffuse Microwave Emission Survey'', 
in the Proceedings from XVI Moriond Astrophysics meeting 
held March March 16-23 in Les Arcs, France, astro-ph/9607100
\bibitem[]{} Kompaneets A.S. 1956, Zh. Eksp. Teor. Fiz., 31, 876 [Sov. Phys. JEPT, 4, 730, (1957)]
\bibitem[]{} Mather J.C., Fixsen D.J., Shafer R.A., Mosier C., Wilkinson, D.T.  1999, ApJ, 512, 511
\bibitem[]{} Nordberg H.P. \& Smoot G.F. 1998, astro-ph/9805123
\bibitem[]{} Pagana E., Villa F., 1996, ``The LOBO Satellite Mission:
Feasibility Study and Preliminary cost evaluation'', Int. Rep. C.I.F.S. - 1996
\bibitem[]{} Press W.H., Teukolsky S.A., Vetterling W.T., Flannery B.P. 1992,
``Numerical Recipes in Fortran'', second edition,
Cambridge University Press, USA
\bibitem[]{} Salvaterra R. \& Burigana C. 2000, Int. Rep. ITeSRE/CNR 270/2000, March, 
astro-ph/0206350
\bibitem[]{} Salvaterra R. \& Burigana C. 2002, MNRAS, 336, 592
\bibitem[]{} Sironi G., Bonelli G., Dall'Oglio G., Pagana E., De~Angeli S., Perelli M., 1995, 
Astroph. Lett. Comm., 32, 31 
\bibitem[]{} Staggs S.T., Jarosik N.C., Meyer S.S., Wilkinson D.T. 1996, ApJ, 473, L1
\label{lastpage}
\end{thebibliography}
\end{document}